\documentclass[preprint,showpacs,showkeys,preprintnumbers,amsmath,amssymb]{revtex4}
\usepackage{graphicx}
\usepackage{dcolumn}
\usepackage{bm}
\def\graphsize{0.5}
\begin{document}
\preprint{}
\title{Analysis of ensemble learning using simple\\
perceptrons based on online learning theory}
\author{Seiji Miyoshi}
\email{miyoshi@kobe-kosen.ac.jp}
\author{Kazuyuki Hara${}^{\dagger}$}
\author{Masato  Okada${}^{\ddagger}$}
\affiliation{
${}^{*}$Department of Electronic Engineering, 
Kobe City College of Technology,
Gakuenhigashi-machi 8-3, Nishi-ku, Kobe, 651-2194 Japan\\
${}^{\dagger}$Department of Electronics and Information Engineering, 
Tokyo Metropolitan College of Technology, 
Higashi-oi 1-10-40, Shinagawa-ku, Tokyo, 140-0011 Japan\\
${}^{\ddagger}$Laboratory for Mathematical Neuroscience, 
RIKEN Brain Science Institute,
Hirosawa 2-1, Wako, Saitama, 351-0198 Japan\\
Intelligent Cooperation and Control, PRESTO, 
Japan Science and Technology Agency,\\
Hirosawa 2-1, Wako, Saitama, 351-0198 Japan
}
\date{\today}


\begin{abstract}
Ensemble learning of $K$ nonlinear perceptrons,
which determine their outputs by sign functions,
is discussed within the framework of online learning
and statistical mechanics.
One purpose of statistical learning theory
is to theoretically obtain the generalization error.
This paper shows that 
ensemble generalization error can be calculated 
by using two order parameters, that is, 
the similarity between a teacher and a student, and
the similarity among students.
The differential equations that describe
the dynamical behaviors of these order parameters
are derived in the case of general learning rules.
The concrete forms of these differential equations are derived
analytically in the cases of three well-known rules:
Hebbian learning, perceptron learning and
AdaTron learning.
Ensemble generalization errors of these three rules
are calculated by using the results determined by 
solving their differential equations.
As a result, these three rules show 
different characteristics in their affinity for ensemble 
learning, that is ``maintaining variety among students." 
Results show that 
AdaTron learning is superior to the other two
rules with respect to that affinity.
\end{abstract}
\maketitle

\section{INTRODUCTION}
Ensemble learning has recently attracted the attention 
of many researchers
\cite{Abe,Breiman,Freund,Hara,Krogh,Urbanczik}.
Ensemble learning means 
to combine many rules or learning machines
(students in the following) that perform poorly.
Theoretical studies analyzing the generalization 
performance by using statistical mechanics\cite{Hertz,Opper} 
have been performed vigorously\cite{Hara,Krogh,Urbanczik}.

Hara and Okada\cite{Hara} theoretically analyzed 
the case in which students are linear perceptrons.
Their analysis was performed with statistical mechanics,
focusing on the fact that the output of a new perceptron, 
whose connection weight is equivalent to 
the mean of those of students,
is identical to the mean outputs of students.
Krogh and Sollich\cite{Krogh} analyzed ensemble learning of 
linear perceptrons with noises within the framework of
batch learning.
They showed that 
the generalization performance can be optimized 
by choosing the best size of learning samples
for a large $K$ limit, where $K$ is the number of students,
and that the generalization performance can be improved 
by dividing learning samples in the noisy situation
when $K$ is finite.

On the other hand, Hebbian learning, perceptron learning
and AdaTron learning are well-known as learning rules 
for a nonlinear perceptron, which decides its output 
by sign function\cite{NishimoriE,Anlauf,Biehl,Inoue}.
Urbanczik\cite{Urbanczik} analyzed ensemble learning 
of nonlinear perceptrons
that decide their outputs by sign functions 
for a large $K$ limit
within the framework of online learning\cite{Saad}.
He treated a generalized learning rule that he termed a
``soft version of perceptron learning," which includes
both Hebbian learning and perceptron learning
as special cases,
and discussed it from the viewpoint of generalization error.
As a result, he showed that 
though an ensemble usually has superior performance to a 
single student, an ensemble has 
no special advantage in the optimized case 
within the framework of the soft version of perceptron learning.
He considered a limit of ensemble learning. 

Though Urbanczik discussed ensemble learning of nonlinear 
perceptrons within the framework of online learning, 
he treated only the case in which 
the number $K$ of students is large enough.
Determining differences among ensemble learnings with Hebbian learning,
perceptron learning and AdaTron learning (three
typical learning rules), is a very attractive problem,
but it is one that has never been analyzed to 
the best of our knowledge.

Based on the past studies, 
we discuss ensemble learning of $K$ nonlinear
perceptrons, which decide their outputs by sign functions
within the framework of online learning and finite $K$
\cite{NC2003-7,JNNS2003}.
First, we show that an ensemble generalization error of 
$K$ students can be calculated by using two order parameters:
one is a similarity between a teacher and a student, 
the other is a similarity among students.
Next, we derive differential equations that describe
dynamical behaviors of these order parameters 
in the case of general learning rules.
After that, we derive concrete differential equations
about three well-known learning rules:
Hebbian learning, perceptron learning and AdaTron 
learning. 
We calculate the ensemble generalization errors 
by using results obtained through solving these equations
numerically.
Two methods are treated to decide an ensemble output.
One is the majority vote of students, 
and the other is an output of a new perceptron 
whose connection weight equals the 
mean of those of students.
As a result, we show that 
these three learning rules have different properties 
with respect to an affinity for ensemble learning, 
and AdaTron learning,
which is known to have the best asymptotic property
\cite{NishimoriE,Anlauf,Biehl,Inoue},
is the best among the three learning rules
within the framework of ensemble learning.

\section{MODEL}
Each student treated in this paper is a perceptron 
that decides its output by a sign function.
An ensemble of $K$ students is considered.
Connection weights of students are
$\mbox{\boldmath $J$}_1,\mbox{\boldmath $J$}_2,\cdots,
\mbox{\boldmath $J$}_K$.
$\mbox{\boldmath $J$}_k=(J_{k1},\cdots,J_{kN}), k=1,2,\cdots,K$
and input $\mbox{\boldmath $x$}=(x_1,\cdots,x_N)$
are $N$ dimensional vectors.
Each component $x_i$ of $\mbox{\boldmath $x$}$
is assumed to be an independent random variable
that obeys the Gaussian distribution ${\cal N}(0,1/N)$.
Each component of $\mbox{\boldmath $J$}_k^0$, that is the initial 
value of $\mbox{\boldmath $J$}_k$, 
is assumed to be generated according to 
the Gaussian distribution ${\cal N}(0,1)$ independently.
Thus,
\begin{equation}
\left\langle x_i\right\rangle=0, \ \left\langle (x_i)^2\right\rangle=\frac{1}{N},
\label{eqn:x}
\end{equation}
\begin{equation}
\left\langle J_{ki}^0\right\rangle=0, \ \left\langle (J_{ki}^0)^2\right\rangle=1,
\end{equation}
where $\langle \cdot \rangle$ denotes the average.
Each student's output is
$\mbox{sgn}(u_1l_1),\mbox{sgn}(u_2l_2),\cdots,\mbox{sgn}(u_Kl_K)$ 
where
\begin{eqnarray}
\mbox{sgn}(ul)
&=&\left\{
\begin{array}{ll}
+1,            & ul \geq 0 , \\
-1,            & ul <    0 ,
\end{array}
\right.\label{eqn:sgn} \\
u_kl_k
&=&\mbox{\boldmath $J$}_k \cdot \mbox{\boldmath $x$}.
\label{eqn:u_k}
\end{eqnarray}
Here, $l_k$ denotes the length of student $\mbox{\boldmath $J$}_k$.
This is one of the order parameters treated in this paper
and will be described in detail later.
In this paper, $u_k$ is called a normalized internal 
potential of a student.

The teacher is also perceptron 
that decides its output by a sign function.
The teacher's connection weight is $\mbox{\boldmath $B$}$.
In this paper, $\mbox{\boldmath $B$}$ is assumed to be fixed
where $\mbox{\boldmath $B$}=(B_1,\cdots,B_N)$ is also an $N$ 
dimensional vector.
Each component $B_i$ is assumed to be generated according to 
the Gaussian distribution ${\cal N}(0,1)$ independently.
Thus,
\begin{equation}
\left\langle B_i\right\rangle=0, \ \left\langle (B_i)^2\right\rangle=1.
\end{equation}
The teacher's output is $\mbox{sgn}(v)$ where
\begin{eqnarray}
v&=&\mbox{\boldmath $B$} \cdot \mbox{\boldmath $x$}.
\label{eqn:v}
\end{eqnarray}
Here, $v$ represents an internal potential of the teacher.
For simplicity, the connection weight of a student and 
that of the teacher 
are simply called student and teacher, respectively.

In this paper the thermodynamic limit $N\rightarrow \infty$
is also treated. Therefore,
\begin{equation}
|\mbox{\boldmath $x$}|=1,\ \ 
|\mbox{\boldmath $B$}|=\sqrt{N},\ \ 
|\mbox{\boldmath $J$}_k^0|=\sqrt{N},
\label{eqn:xBJ}
\end{equation}
where $|\cdot|$ denotes a vector norm.
Generally, a norm of student $|\mbox{\boldmath $J$}_k|$
changes as the time step proceeds.
Therefore, the ratio $l_k$ of the norm to $\sqrt{N}$
is considered and 
is called a length of student $\mbox{\boldmath $J$}_k$.
That is, 
\begin{equation}
|\mbox{\boldmath $J$}_k|=l_k\sqrt{N},
\label{eqn:l}
\end{equation}
where $l_k$ is one of the order parameters treated in this paper.

The common input $\mbox{\boldmath $x$}$ is presented 
to the teacher and all students in the same order.
Each student compares its output and 
an output of the teacher for input $\mbox{\boldmath $x$}$.
Each student's connection weight is corrected
for the increasing probability that 
the student output agrees with that of the teacher.
This procedure is called learning,
and a method of learning is called learning rule,
of which Hebbian learning, perceptron learning and AdaTron learning
are well-known examples\cite{NishimoriE,Anlauf,Biehl,Inoue}.
Within the framework of online learning,
information that can be used for correction 
other than that regarding a student itself
is only input $\mbox{\boldmath $x$}$ and an output of
the teacher for that input. 
Therefore, the update can be expressed as follows,
\begin{eqnarray}
\mbox{\boldmath $J$}^{m+1}_k
&=& \mbox{\boldmath $J$}^m_k+f^m_k\mbox{\boldmath $x$}^m,
\label{eqn:general1} \\
f^m_k
&=& f(\mbox{sgn}(v^m),u^m_k),
\label{eqn:general2}
\end{eqnarray}
where $m$ denotes time step, and 
$f$ is a function determined by learning rule.

In this paper, two
methods are treated to determine an ensemble output.
One is the majority vote of $K$ students,
which means 
an ensemble output is decided to be $+1$
if students whose outputs are $+1$ exceed 
the number of students whose outputs are $-1$, 
and $-1$ in the opposite case.

Another method for deciding an ensemble output
is adopting an output of a new perceptron 
whose connection weight is the mean of the weights of 
$K$ students.
This method is simply called the weight mean in this paper.

\section{THEORY} \label{sec:ge}
In this paper, the majority vote and the weight mean
are treated to determine an ensemble output.
We use 
\begin{equation}
\epsilon 
= \Theta \left(
-\mbox{sgn}\left(\mbox{\boldmath $B$} \cdot \mbox{\boldmath $x$}\right)
\mbox{sgn}\left(
\sum_{k=1}^K\mbox{sgn}\left(\mbox{\boldmath $J$}_k \cdot \mbox{\boldmath $x$}
\right)\right)\right), \label{eqn:epsilonMV}
\end{equation}
and 
\begin{equation}
\epsilon 
= \Theta \left(
-\mbox{sgn}\left(\mbox{\boldmath $B$} \cdot \mbox{\boldmath $x$}\right)
\mbox{sgn}\left(\left(
\frac{1}{K}
\sum_{k=1}^K \mbox{\boldmath $J$}_k\right) 
\cdot \mbox{\boldmath $x$}\right)\right), \label{eqn:epsilonWM}
\end{equation}
as error $\epsilon$ for the 
majority vote and the weight mean, respectively.
Here, $\mbox{\boldmath $\epsilon$}$,
$\mbox{\boldmath $x$}$ and
$\mbox{\boldmath $J$}_k$
denote
$\mbox{\boldmath $\epsilon$}^m$,
$\mbox{\boldmath $x$}^m$ and
$\mbox{\boldmath $J$}_k^m$,
respectively.
However, superscripts $m$, which represent time steps, are 
omitted for simplicity.
Then, $\Theta(\cdot)$ is the step function defined as
\begin{eqnarray}
\Theta (z)
&=&\left\{
\begin{array}{ll}
+1,           & z \geq 0 , \\
0,            & z <    0 .
\end{array}
\right.\label{eqn:Theta}
\end{eqnarray}

In both cases, $\epsilon=0$ if an ensemble output agrees 
with that of the teacher and $\epsilon=1$ otherwise.
Generalization error $\epsilon_g$
is defined as the average of error $\epsilon$
over the probability distribution $p(\mbox{\boldmath $x$})$
of input $\mbox{\boldmath $x$}$.
The generalization error $\epsilon_g$
can be regarded as the probability 
that an ensemble output disagrees with that of the teacher
for a new input $\mbox{\boldmath $x$}$.
One purpose of statistical learning theory
is to theoretically obtain generalization error.
In the case of a majority vote,
using Eqs. (\ref{eqn:u_k}), (\ref{eqn:v}) 
and (\ref{eqn:epsilonMV}), we obtain
\begin{equation}
\epsilon 
=  \Theta \left(
-\mbox{sgn}(v)
\sum_{k=1}^K\mbox{sgn}\left(u_k\right)\right).
\end{equation}
In the case of a weight mean,
using Eqs. (\ref{eqn:u_k}), (\ref{eqn:v}) and (\ref{eqn:epsilonWM}),
we obtain
\begin{equation}
\epsilon 
= \Theta \left(
-\mbox{sgn}\left(v\right)
\mbox{sgn}\left(
\sum_{k=1}^K u_k\right)\right).
\end{equation}
That is error $\epsilon$ 
can be described as $\epsilon=\epsilon(\{u_k\},v)$
by using a normalized internal potential $u_k$ for the student
and an internal potential $v$ for the teacher in both cases.
Therefore, the generalization error $\epsilon_g$ can be also
described as 
\begin{eqnarray}
\epsilon_g 
&=& \int d\mbox{\boldmath $x$} p(\mbox{\boldmath $x$})
\epsilon
\nonumber \\
&=& \int \prod_{k=1}^Kdu_kdv 
      p(\{u_k\},v)
      \epsilon(\{u_k\},v),
\label{eqn:eg}
\end{eqnarray}
by using the probability distribution $p(\{u_k\},v)$
of $u_k$ and $v$.
From Eq. (\ref{eqn:u_k}), we can write
\begin{equation}
u_k=\frac{1}{l_k}\sum_{i=1}^N J_{ki}x_i,
\end{equation}
where $J_{ki}x_i, i=1,\cdots,N$
are independent and identically distributed 
random variables.
In the same manner, 
from Eq. (\ref{eqn:v}), we can write
\begin{equation}
v=\sum_{i=1}^N B_ix_i,
\end{equation}
where $B_ix_i, i=1,\cdots,N$
are independent and identically distributed 
random variables.
Since the thermodynamic limit $N\rightarrow \infty$
is also considered in this paper, $u_k$ and $v$ obey
the multiple Gaussian distribution
based on the central limit theorem.
The discussion in this paper falls within the framework of
online learning, which means input $\mbox{\boldmath $x$}$,
once used for an update, is abandoned 
and $\mbox{\boldmath $x$}$ for each time step 
is generated according to the Gaussian distribution
of Eq. (\ref{eqn:x}). 
Therefore, since an input $\mbox{\boldmath $x$}$ and 
a student $\mbox{\boldmath $J$}_k$ 
have no correlation with each other,
from Eq. (\ref{eqn:u_k}), the mean and the variance of $u_k$
are 
\begin{eqnarray}
\left\langle u_k \right\rangle
&=& \left\langle \frac{1}{l_k}
    \mbox{\boldmath $J$}_k \cdot \mbox{\boldmath $x$}\right\rangle \\
&=& \left\langle \frac{1}{l_k}
    \sum_{i=1}^N J_{ki}x_i \right\rangle\\
&=& \frac{1}{l_k}
    \sum_{i=1}^N \left\langle J_{ki}\right\rangle \left\langle x_i\right\rangle \\
&=& 0, \label{eqn:mean_u_k} \\
\left\langle (u_k)^2 \right\rangle
&=& \left\langle \left(\frac{1}{l_k}
    \mbox{\boldmath $J$}_k \cdot \mbox{\boldmath $x$}\right)^2\right\rangle \\
&=& \left\langle \frac{1}{l_k^2}
    \sum_{i=1}^N J_{ki}x_i \sum_{j=1}^N J_{kj}x_j \right\rangle\\
&=& \frac{1}{l_k^2}
    \sum_{i=1}^N \left\langle (J_{ki})^2\right\rangle \left\langle (x_i)^2\right\rangle \\
&=& 1, \label{eqn:var_u_k} 
\end{eqnarray}
respectively.
In the same manner, 
since an input $\mbox{\boldmath $x$}$ and a teacher
$\mbox{\boldmath $B$}$ have no correlation with each other,
from Eq. (\ref{eqn:v}), the mean and the variance of $v$
are
\begin{eqnarray}
\left\langle v \right\rangle
&=& \left\langle 
    \mbox{\boldmath $B$} \cdot \mbox{\boldmath $x$}\right\rangle \\
&=& \left\langle 
    \sum_{i=1}^N B_ix_i \right\rangle\\
&=& \sum_{i=1}^N \left\langle B_i\right\rangle \left\langle x_i\right\rangle \\
&=& 0, \label{eqn:mean_v} \\
\left\langle v^2 \right\rangle
&=& \left\langle \left(
    \mbox{\boldmath $B$} \cdot \mbox{\boldmath $x$}\right)^2\right\rangle \\
&=& \left\langle
    \sum_{i=1}^N B_ix_i \sum_{j=1}^N B_jx_j \right\rangle\\
&=& \sum_{i=1}^N \left\langle (B_i)^2\right\rangle \left\langle (x_i)^2\right\rangle \\
&=& 1, \label{eqn:var_v} 
\end{eqnarray}
respectively.

From these, all diagonal components of the covariance matrix
$\mbox{\boldmath $\Sigma$}$ of $p(\{u_k\},v)$ equal unity.

Let us discuss a direction cosine between connection weights
as preparation for obtaining non-diagonal components.
First, $R_k$ is defined as a direction cosine between 
a teacher $\mbox{\boldmath $B$}$ and a student $\mbox{\boldmath $J$}_k$.
That is,
\begin{equation}
R_k\equiv \frac{\mbox{\boldmath $B$}\cdot\mbox{\boldmath $J$}_k}
{|\mbox{\boldmath $B$}||\mbox{\boldmath $J$}_k|}
=\frac{1}{l_kN}\sum_{i=1}^NB_iJ_{ki}.
\end{equation}
When a teacher $\mbox{\boldmath $B$}$ and 
a student $\mbox{\boldmath $J$}_k$ have no correlation,
$R_k=0$, and 
$R_k=1$ when the 
directions of $\mbox{\boldmath $B$}$ and $\mbox{\boldmath $J$}_k$
agree. 
Therefore, $R_k$ is called the similarity  (overlap in other word)
between teacher and student 
in the following.
Furthermore, $R_k$ is the second order parameter treated in this paper.
Next, $q_{kk'}$ is defined as a direction cosine between 
a student $\mbox{\boldmath $J$}_k$ and 
another student $\mbox{\boldmath $J$}_{k'}$.
That is,
\begin{equation}
q_{kk'}\equiv
\frac{\mbox{\boldmath $J$}_k\cdot\mbox{\boldmath $J$}_{k'}}
{|\mbox{\boldmath $J$}_k||\mbox{\boldmath $J$}_{k'}|}
=\frac{1}{l_kl_{k'}N}\sum_{i=1}^NJ_{ki}J_{k'i},
\label{eqn:q}
\end{equation}
where $k\neq k'$.
When a student $\mbox{\boldmath $J$}_k$ and 
another student $\mbox{\boldmath $J$}_{k'}$ have no correlation,
$q_{kk'}=0$, and $q_{kk'}=1$ when the directions of $\mbox{\boldmath $J$}_k$ 
and $\mbox{\boldmath $J$}_{k'}$ agree. 
Therefore, $q_{kk'}$ is called the 
similarity among students in the following,
and $q_{kk'}$ is the third order parameter treated in this paper.

Covariance between an internal potential $v$ of 
a teacher $\mbox{\boldmath $B$}$ and
a normalized internal potential $u_k$ of
a student $\mbox{\boldmath $J$}_k$
equals a similarity $R_k$ between
a teacher $\mbox{\boldmath $B$}$ and
a student $\mbox{\boldmath $J$}_k$ as follows,
\begin{eqnarray}
\left\langle vu_k \right\rangle
&=& \left\langle \frac{1}{l_k}\sum_{i=1}^N B_ix_i \sum_{j=1}^N J_{kj}x_j \right\rangle \\
&=& \frac{1}{l_k}\sum_{i=1}^N\left\langle B_iJ_{ki}\right\rangle \left\langle (x_i)^2\right\rangle \\
&=& \frac{1}{l_kN}\sum_{i=1}^N\left\langle B_iJ_{ki}\right\rangle \\
&=& R_k.
\end{eqnarray}
Covariance between a normalized internal potential $u_k$ of 
a student $\mbox{\boldmath $J$}_k$ and
a normalized internal potential $u_{k'}$ of
another student $\mbox{\boldmath $J$}_{k'}$
equals a similarity $q_{kk'}$ among
students as follows,
\begin{eqnarray}
\left\langle u_ku_{k'} \right\rangle
&=& \left\langle \frac{1}{l_kl_{k'}}
    \sum_{i=1}^N J_{ki}x_i \sum_{j=1}^N J_{k'j}x_j \right\rangle \\
&=& \frac{1}{l_kl_{k'}}
    \sum_{i=1}^N\left\langle J_{ki}J_{k'i}\right\rangle \left\langle (x_i)^2\right\rangle \\
&=& \frac{1}{l_kl_{k'}N}\sum_{i=1}^N\left\langle J_{ki}J_{k'i}\right\rangle \\
&=& q_{kk'}.
\end{eqnarray}
Therefore, Eq. (\ref{eqn:eg}) can be rewritten as
\begin{eqnarray}
\epsilon_g
&=& \int \prod_{k=1}^Kdu_kdv 
      p(\{u_k\},v)
      \epsilon(\{u_k\},v), \label{eqn:eg2}\\
p(\{u_k\},v)
&=& \frac{1}{(2\pi)^\frac{K+1}{2}|
    \mbox{\boldmath $\Sigma$}|^\frac{1}{2}}\nonumber \\
&\times&
    \exp\left(-\frac{(\{u_k\},v)
    \mbox{\boldmath $\Sigma$}^{-1}(\{u_k\},v)^T}{2}\right),
\nonumber \\
& &    \label{eqn:P} \\
\mbox{\boldmath $\Sigma$}
  &=&
   \left(
   \arraycolsep=3pt
   \begin{array}{ccccc}
     \!1\!    & \!q_{12}\! & \!\ldots\! & \!q_{1K}\! & \!R_1\!    \\
     \!q_{21}\! & \!1\!      & \!\ddots\! & \!\vdots\! & \!\vdots\! \\
     \!\vdots\! & \!\ddots\! & \!\ddots\! & \!q_{K-1,K}\! & \!\vdots\! \\
     \!q_{K1}\! & \!\ldots\! & \!q_{K,K-1}\! & \!1\!      & \!R_K\!    \\
     \!R_1\!    & \!\ldots\! & \!\ldots\! & \!R_K\!     & \!1\!
   \end{array}
   \right).\nonumber \\
& & \label{eqn:Sigma_for_eg}
\end{eqnarray}
As a result, a generalization error $\epsilon_g$
can be calculated if all similarities $R_k$ and $q_{kk'}$
are obtained.
Let us thus discuss differential equations that describe
dynamical behaviors of these order parameters.
In this paper, norms of inputs, teacher and students 
are set as Eq. (\ref{eqn:xBJ});
influence of input can be replaced with the average over
the distribution of inputs (sample average) 
in a large $N$ limit.
This idea is called self-averaging in statistical mechanics.
Differential equations regarding $l_k$ and $R_k$
for general learning rules
have been obtained based on self-averaging
as follows\cite{NishimoriE},
\begin{eqnarray}
\frac{dl_k}{dt}&=&
\left\langle f_k u_k\right\rangle +\frac{\left\langle f_k^2\right\rangle }{2l_k}
\label{eqn:dldt}, \\
\frac{dR_k}{dt}&=&
\frac{\left\langle f_kv\right\rangle -\left\langle f_k u_k\right\rangle R_k}{l_k}
- \frac{R_k}{2l_k^2}\left\langle f_k^2\right\rangle ,
\label{eqn:dRdt}
\end{eqnarray}
where $\left\langle \cdot \right\rangle$ stands for 
the sample average.
That is,
\begin{eqnarray}
\left\langle f_k u_k\right\rangle
&=& \int du_kdvp_2(u_k,v)f(\mbox{sgn}(v),u_k)u_k, \nonumber \\
& & \label{eqn:fkuk}\\
\left\langle f_k v\right\rangle
&=& \int du_kdvp_2(u_k,v)f(\mbox{sgn}(v),u_k)v, \nonumber \\
& &  \label{eqn:fkv}\\
\left\langle f_k^2\right\rangle
&=& \int dudvp_2(u_k,v)\left(f(\mbox{sgn}(v),u_k)\right)^2, \nonumber \\
& &   \label{eqn:fk^2}\\
p_2(u_k,v)
&=& \frac{1}{2\pi|\mbox{\boldmath $\Sigma$}_2|^\frac{1}{2}}\nonumber \\
&\times&    \exp\left(-\frac{(u_k,v)
    \mbox{\boldmath $\Sigma$}_2^{-1}(u_k,v)^T}{2}\right),
    \label{eqn:P2} \\
\mbox{\boldmath $\Sigma$}_2
&=&
   \left(
   \arraycolsep=3pt
   \begin{array}{cc}
     1      & R_k    \\
     R_k    & 1
   \end{array}
   \right). \label{eqn:Sigma2}
\end{eqnarray}

Next, let us derive a differential equation 
regarding $q_{kk'}$ for the general learning rule.
Considering a student $\mbox{\boldmath $J$}_k$ and
another student $\mbox{\boldmath $J$}_{k'}$
and rewriting as
$l_k^m\rightarrow l_k$,
$l_k^{m+1}\rightarrow l_k+dl_k$,
$q_{kk'}^m\rightarrow q_{kk'}$,
$q_{kk'}^{m+1}\rightarrow q_{kk'}+dq_{kk'}$ and
$1/N\rightarrow dt$,
a differential equation regarding $q$ is obtained
as follows\cite{Hara},
\begin{eqnarray}
\frac{dq_{kk'}}{dt}
&=&
\frac{\left\langle f_{k'} u_k\right\rangle -q_{kk'}\left\langle f_{k'} u_{k'}\right\rangle}{l_{k'}}\nonumber \\
&+& \frac{\left\langle f_k u_{k'}\right\rangle -q_{kk'}\left\langle f_k u_k\right\rangle}{l_k}
\nonumber \\
&+& \frac{\left\langle f_kf_{k'}\right\rangle}{l_kl_{k'}}
-\frac{q_{kk'}}{2}
\left(\frac{\left\langle f_k^2\right\rangle}{l_k^2}
+\frac{\left\langle f_{k'}^2\right\rangle}{l_{k'}^2}\right),
\label{eqn:dqdt}
\end{eqnarray}
from Eqs. (\ref{eqn:general1}), (\ref{eqn:q}), (\ref{eqn:dldt})
and self-averaging,
where
\begin{eqnarray}
\left\langle f_{k} u_{k'}\right\rangle
&=& \int du_kdu_{k'}dvp_3(u_k,u_{k'},v)\nonumber \\
& & \times f(\mbox{sgn}(v),u_{k})u_{k'},
\label{eqn:fkuk'}\\
\left\langle f_{k'} u_k\right\rangle
&=& \int du_kdu_{k'}dvp_3(u_k,u_{k'},v)\nonumber \\
& & \times f(\mbox{sgn}(v),u_{k'})u_k,
\label{eqn:fk'uk}\\
\left\langle f_k f_{k'}\right\rangle
&=& \int du_kdu_{k'}dvp_3(u_k,u_{k'},v)\nonumber \\
& & \times f(\mbox{sgn}(v),u_{k})f(\mbox{sgn}(v),u_{k'}),
 \label{eqn:fkfk'} \\
p_3(u_k,u_{k'},v)
&=& \frac{1}{(2\pi)^\frac{3}{2}|
    \mbox{\boldmath $\Sigma$}_3|^\frac{1}{2}}\nonumber \\
&\times& 
    \exp\left(-\frac{(u_k,u_{k'},v)
    \mbox{\boldmath $\Sigma$}_3^{-1}(u_k,u_{k'},v)^T}{2}\right),
\nonumber \\
& &     \label{eqn:P3} \\
\mbox{\boldmath $\Sigma$}_3
  &=&
   \left(
   \arraycolsep=3pt
   \begin{array}{ccc}
     1       & q_{kk'} & R_k    \\
     q_{k'k} & 1       & R_{k'} \\
     R_k     & R_{k'}  & 1
   \end{array}
   \right).\label{eqn:Sigma3}
\end{eqnarray}

\section{RESULT}
\subsection{Conditions of analytical calculations}
As described above,
in this paper each component of 
initial value $\mbox{\boldmath $J$}_k^0$ 
of student $\mbox{\boldmath $J$}_k$ 
and teacher $\mbox{\boldmath $B$}$
is generated independently 
according to 
the Gaussian distribution ${\cal N}(0,1)$,
and the thermodynamic limit $N\rightarrow \infty$
is considered.
Therefore, all $\mbox{\boldmath $J$}_k^0$ and $\mbox{\boldmath $B$}$
are orthogonal to each other. That is,
\begin{equation}
R_k^0=0,\ \ \ q_{kk'}^0=0.
\label{eqn:Rqinit}
\end{equation}
From Eq. (\ref{eqn:Rqinit}) and symmetry of students,
we can write
\begin{equation}
\left\langle f_k u_{k'}\right\rangle =\left\langle f_{k'} u_k\right\rangle,
\ \ \ 
\left\langle f_k f_{k'}\right\rangle =\left\langle f_{k'} f_k\right\rangle
\label{eqn:fkuk'fkfk'}
\end{equation}
in Eq. (\ref{eqn:dqdt}).
From Eq. (\ref{eqn:Rqinit}) and symmetry among students,
we omit subscripts $k,k'$ from order parameters
$l_k, R_k$ and $q_{kk'}$
in Eqs. (\ref{eqn:dldt})--(\ref{eqn:Sigma3})
and write them as $l, R$ and $q$.
In the following sections, 
we analytically obtain five sample averages
$\left\langle f_k u_k\right\rangle $,
$\left\langle f_k v\right\rangle $,
$\left\langle f_k^2\right\rangle $,
$\left\langle f_k u_{k'}\right\rangle $ and 
$\left\langle f_k f_{k'}\right\rangle $
concretely,
which are necessary to solve
Eqs. (\ref{eqn:dldt})--(\ref{eqn:Sigma3})
with respect to typical learning rules
under the conditions 
given in Eqs. (\ref{eqn:Rqinit})--(\ref{eqn:fkuk'fkfk'}).
$R$ and $q$ are obtained by solving the above sample averages 
and Eqs. (\ref{eqn:dldt})，(\ref{eqn:dRdt})，
(\ref{eqn:dqdt}) and (\ref{eqn:Rqinit}) numerically.
We obtain numerical ensemble generalization errors $\epsilon_g$
by solving Eq. (\ref{eqn:eg2}) with the obtained $R$ and $q$.

\subsection{Hebbian learning}
The update procedure for Hebbian learning is
\begin{equation}
f(\mbox{sgn}(v),u)=\mbox{sgn}(v).
\label{eqn:Hebb}
\end{equation}

Using this expression,
$\left\langle f_k u_k\right\rangle $,
$\left\langle f_k v\right\rangle $ and
$\left\langle f_k^2\right\rangle $
in the case of Hebbian learning can be obtained as follows
by executing Eqs.(\ref{eqn:fkuk})--(\ref{eqn:fk^2})
analytically\cite{NishimoriE,Engel}.
\begin{equation}
\left\langle f_k u_k\right\rangle = \frac{2R}{\sqrt{2\pi}}, \ \ 
\left\langle f_k v\right\rangle = \sqrt{\frac{2}{\pi}}, \ \ 
\left\langle f_k^2\right\rangle =1 \label{eqn:fufvf2H}.
\end{equation}

In this section, $\left\langle f_{k} u_{k'}\right\rangle $ and
$\left\langle f_{k} f_{k'}\right\rangle $ are derived.
Since Eq.(\ref{eqn:Hebb}) is independent of $u$, we obtain
\begin{eqnarray}
\left\langle f_{k} u_{k'}\right\rangle 
&=& \left\langle f_k u_k\right\rangle 
= \frac{2R}{\sqrt{2\pi}}, \label{eqn:faubH} \\
\left\langle f_{k} f_{k'}\right\rangle 
&=& \left\langle (\mbox{sgn}(v))^2\right\rangle 
= 1.
\label{eqn:fafbH}
\end{eqnarray}

\begin{figure}[htbp]
\begin{center}
\includegraphics[width=\graphsize\linewidth,keepaspectratio]{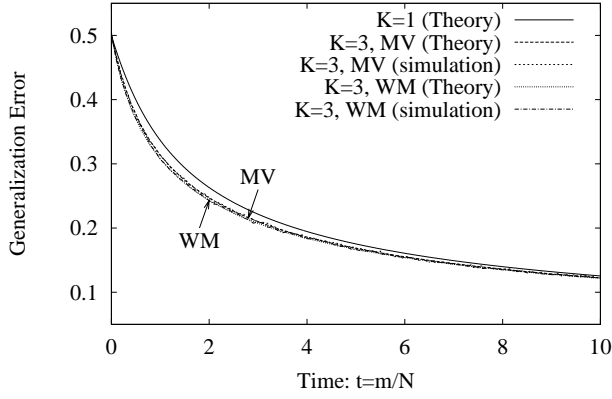}
\caption{Dynamical behaviors of ensemble generalization error $\epsilon_g$
in Hebbian learning.}
\label{fig:Hebb}
\end{center}
\end{figure}

$R$ and $q$ have been obtained by solving 
Eqs. (\ref{eqn:dldt}), (\ref{eqn:dRdt}), (\ref{eqn:dqdt}), 
(\ref{eqn:Rqinit}), (\ref{eqn:fkuk'fkfk'}), 
(\ref{eqn:fufvf2H})--(\ref{eqn:fafbH}) numerically.
We have obtained numerical 
ensemble generalization errors $\epsilon_g$
in the case of $K=3$ by using 
Eqs. (\ref{eqn:eg2})--(\ref{eqn:Sigma_for_eg})
and the above $R$ and $q$.
Figure \ref{fig:Hebb} shows the results.
In this figure, MV and WM indicate the majority vote and the
weight mean, respectively.
Numerical integrations of Eq. (\ref{eqn:eg2})
in theoretical calculations
have been executed by using 
the six-point closed Newton-Cotes formula.
In the computer simulation, $N=10^4$ and ensemble generalization
errors have been obtained through tests using $10^5$ random inputs
at each time step.
In this figure, the result of theoretical calculations of $K=1$
is also shown to clarify the effect of the ensemble.
This figure shows that 
the ensemble generalization errors obtained 
by theoretical calculation
explain the computer simulation quantitatively.

\begin{figure}[htbp]
\begin{center}
\includegraphics[width=\graphsize\linewidth,keepaspectratio]{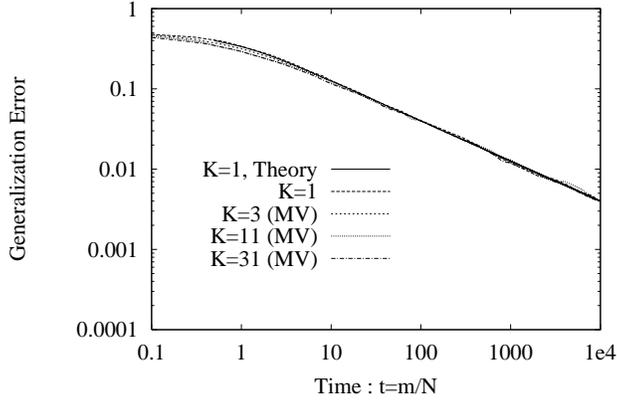}
\caption{Asymptotic behavior of generalization error
of majority vote in Hebbian learning.
Computer simulations, except for the solid line.
Asymptotic order of ensemble learning is the same as
that at $K=1$.}
\label{fig:Hasysim}
\end{center}
\end{figure}

\begin{figure}[htbp]
\begin{center}
\includegraphics[width=\graphsize\linewidth,keepaspectratio]{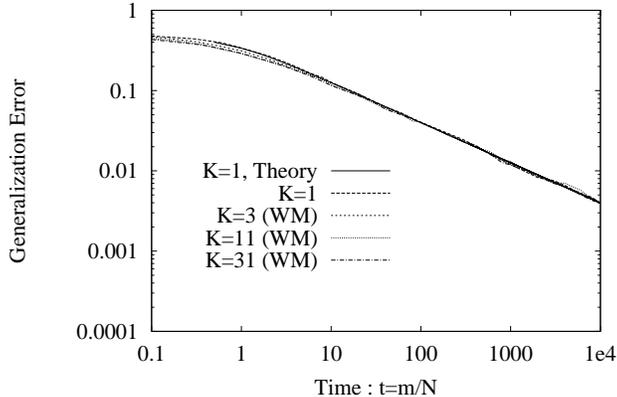}
\caption{Asymptotic behavior of generalization error
of weight mean in Hebbian learning.
Computer simulations, except for the solid line.
Asymptotic order of ensemble learning is the same as
that at $K=1$.}
\label{fig:HasysimWM}
\end{center}
\end{figure}

Figures \ref{fig:Hasysim}--\ref{fig:HasysimWM}
show the results of computer simulations 
where $N=10^3$, $K=1,3,11,31$
until $t=10^4$ in order to investigate asymptotic behaviors
of generalization errors.
Asymptotic behavior of generalization error
in Hebbian learning
in the case of the number $K$ of students at unity
is $O(t^{-\frac{1}{2}})$\cite{NishimoriE}.
Asymptotic orders of the generalization error
in the case of ensemble learning are considered equal to
those of $K=1$, since properties of $K=3,11,31$ are
parallel to those of $K=1$ in these figures.

To clarify the relationship between
$K$ and the effect of ensemble, 
we have obtained theoretical ensemble generalization
errors for various values of $K$. 
Here, it is difficult to
execute numerical integration of Eq. (\ref{eqn:eg2})
when $K>3$ by the Newton-Cotes formula
used in the calculations for
Figure \ref{fig:Hebb}.
Therefore, the Metropolis method, which is a type of 
MonteCarlo method, has been used.
We then orthogonalized the variables of integration
to eliminate the calculation of inverse matrices
of Eq. (\ref{eqn:Sigma_for_eg}).
That is, 
\begin{equation}
u_k=a\bar{u}_k+b\hat{u}+cv, \ \ k=1,2,\cdots,K,
\label{eqn:ortho}
\end{equation}
where $u_k,\bar{u}_k,\hat{u}$ and $v$ obey the Gaussian
distribution ${\cal N}(0,1)$ and 
$\bar{u}_k,\hat{u}$ and $v$ have no correlation with each other.
Considering that subscripts $k,k'$ have been omitted
from order parameters $R_k, q_{kk'}$ and
Eq. (\ref{eqn:Sigma_for_eg}),
conditions that $a,b$ and $c$ must satisfy are
\begin{eqnarray}
a^2+b^2+c^2 &=& 1, \\
b^2+c^2     &=& q, \\
c           &=& R.
\end{eqnarray}
Therefore, 
\begin{eqnarray}
a&=&\sqrt{1-q}, \label{eqn:a}\\
b&=&\sqrt{q-R^2}, \\
c&=&R.
\end{eqnarray}

By using these $a,b$ and $c$, we can rewrite
Eqs. (\ref{eqn:eg2})--(\ref{eqn:Sigma_for_eg})
as follows:
\begin{eqnarray}
\epsilon_g
&=& \int \prod_{k=1}^K d\bar{u}_kp_1(\bar{u}_{k})
      d\hat{u}p_1(\hat{u})dvp_1(v)
      \epsilon(\{a\bar{u}_k+b\hat{u}+cv\},v), \label{eqn:eg3}\\
p_1(u)
&=& \frac{1}{(2\pi)^\frac{1}{2}}
    \exp\left(-\frac{u^2}{2}\right). \label{eqn:P_1}
\end{eqnarray}

These operations orthogonalized the variables of integration
in exchange for
their number having been increased from
$K+1$ to $K+2$.
The multiple Gaussian distribution function $p(\{u_k\},v)$
can be rewritten as products of simple Gaussian 
distribution functions $p_1(\cdot)$ by this orthogonalization.
Thus, calculations of inverse matrices of 
Eq. (\ref{eqn:Sigma_for_eg}) become unnecessary.
These facts have made it easy to perform
the numerical calculations of the generalization 
error for a large $K$.

\begin{figure}[htbp]
\begin{center}
\includegraphics[width=\graphsize\linewidth,keepaspectratio]{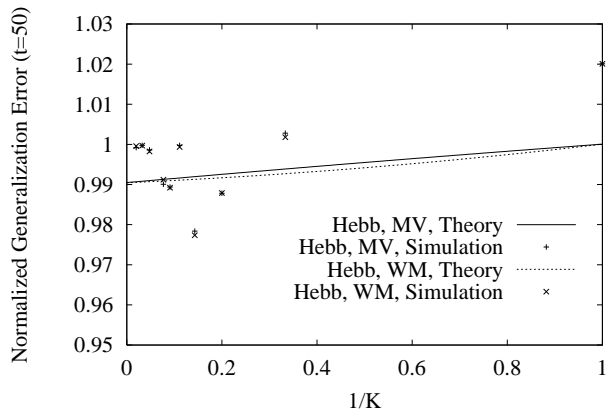}
\caption{Relationship between $K$ and effect of 
ensemble in Hebbian learning.
Ensemble generalization error $\epsilon_g$ for a large $K$ limit
is about 0.99 times 
that of $K=1$.}
\label{fig:Hasy}
\end{center}
\end{figure}

Figure \ref{fig:Hasy} shows the results
obtained by the Metropolis method 
using the values of $R$ and $q$ calculated
numerically for Hebbian learning
and Eqs. (\ref{eqn:a})--(\ref{eqn:P_1}).
Calculations have been executed for 
$K=1$, $3$, $5$, $7$, $9$, $11$, $13$, $21$, $31$ and $51$
in both the majority vote (MV) 
and the weight mean (WM). The number of MonteCarlo steps
is $10^9$. These theoretical results are fitted to 
two quadratic curves.
In this figure, the results of computer simulations
where $N=10^4$，$K=1$, $3$, $5$, $7$, $9$, $11$, 
$13$, $21$, $31$ and $51$ have also been drawn
for comparison with the theoretical calculations.
In the computer simulations, ensemble generalization errors
have been obtained through tests using $10^6$ random inputs.
The figures show the values of $t=50$ for
both theoretical calculations and computer simulations,
and this is the time for which is considered 
that the learnings are sufficiently within the asymptotic regions
with respect to Figures \ref{fig:Hasysim}--\ref{fig:HasysimWM}.
Here, since the relationship between $1/K$ and
ensemble generalization errors shows a straight line
\cite{Hara} in the case of linear perceptrons, 
the abscissa is $1/K$ in 
Figure \ref{fig:Hasy}.
The ordinates have been normalized by
the theoretical ensemble generalization error of $K=1$ and $t=50$.

\subsection{Perceptron learning}
The update procedure for perceptron learning is
\begin{equation}
f(\mbox{sgn}(v),u)=\Theta\left(-uv\right)\mbox{sgn}(v).
\label{eqn:Perceptron}
\end{equation}

Using this expression,
$\left\langle f_k u_k\right\rangle $,
$\left\langle f_k v\right\rangle $ and
$\left\langle f_k^2\right\rangle $
in the case of perceptron learning can be obtained as follows
by executing Eqs.(\ref{eqn:fkuk})--(\ref{eqn:fk^2})
analytically\cite{NishimoriE,Engel}.
\begin{eqnarray}
\left\langle f_k u_k\right\rangle  &=& \frac{R-1}{\sqrt{2\pi}},\ \ \ \ 
\left\langle f_k v\right\rangle = \frac{1-R}{\sqrt{2\pi}}, \label{eqn:fufvP}\\
\left\langle f_k^2\right\rangle &=&
2\int_0^\infty DvH\left(\frac{Rv}{\sqrt{1\!-\!R^2}}\right)\nonumber \\
&=&\frac{1}{\pi}\tan^{\!-\!1}\frac{\sqrt{1\!-\!R^2}}{R}. \label{eqn:f2Ptan}
\end{eqnarray}

In this section, $\left\langle f_{k} u_{k'}\right\rangle $ and
$\left\langle f_{k} f_{k'}\right\rangle $ are derived.
Using Eq. (\ref{eqn:Perceptron}),
$\left\langle f_{k} u_{k'}\right\rangle $ and 
$\left\langle f_{k} f_{k'}\right\rangle $
in the case of perceptron learning
are obtained as follows by executing Eqs. (\ref{eqn:fkuk'}) and
(\ref{eqn:fkfk'}) analytically.
\begin{eqnarray}
\left\langle f_{k} u_{k'}\right\rangle 
&=&\int du_{k} du_{k'} dv p_3(u_{k},u_{k'},v)\nonumber \\
& & \times \Theta(-u_{k} v)\mbox{sgn}(v)u_{k'} \nonumber \\
&=& \frac{R-q}{\sqrt{2\pi}} \label{eqn:faubP} \\
\left\langle f_{k} f_{k'}\right\rangle
&=& \int du_ku_{k'}dvp_3(u_k,u_{k'},v)\nonumber \\
& & \times \Theta(-u_{k} v)\Theta(-u_{k'} v) \nonumber \\
&=& 2\int_0^\infty Dv\int_{\frac{Rv}{\sqrt{1-R^2}}}^\infty Dx
     H\left(z\right) \label{eqn:fafbP}
\end{eqnarray}
where
\begin{eqnarray}
z &\equiv& \frac{-(q-R^2)x+R\sqrt{1-R^2}v}
         {\sqrt{(1-q)(1+q-2R^2)}}
\label{eqn:z}
\end{eqnarray}
and the definitions of $H(u)$ and $Dx$ are
\begin{eqnarray}
H(u)&\equiv&\int_u^\infty Dx \label{eqn:Hu}\\
Dx&\equiv& \frac{dx}{\sqrt{2\pi}}\exp\left(-\frac{x^2}{2}\right).
\label{eqn:Dx}
\end{eqnarray}

\begin{figure}[htbp]
\begin{center}
\includegraphics[width=\graphsize\linewidth,keepaspectratio]{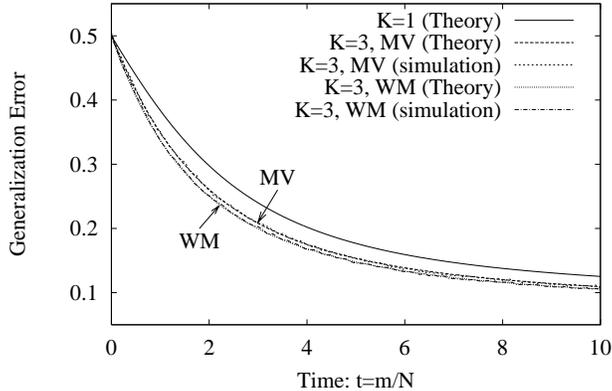}
\caption{Dynamical behaviors of ensemble generalization error $\epsilon_g$
in perceptron learning.}
\label{fig:Perceptron}
\end{center}
\end{figure}

In the same manner as Hebbian learning,
$R$ and $q$ have been obtained by solving 
Eqs. (\ref{eqn:dldt}), (\ref{eqn:dRdt}), (\ref{eqn:dqdt}), 
(\ref{eqn:Rqinit}), (\ref{eqn:fkuk'fkfk'}), 
(\ref{eqn:fufvP})--(\ref{eqn:fafbP}) numerically.
We have obtained numerical 
ensemble generalization errors $\epsilon_g$
in the case of $K=3$ by using 
Eqs. (\ref{eqn:eg2})--(\ref{eqn:Sigma_for_eg})
and the above $R$ and $q$.
Figure \ref{fig:Perceptron} shows the results.
This figure shows that 
the ensemble generalization errors obtained 
by theoretical calculation
explain the computer simulation quantitatively.

\begin{figure}[htbp]
\begin{center}
\includegraphics[width=\graphsize\linewidth,keepaspectratio]{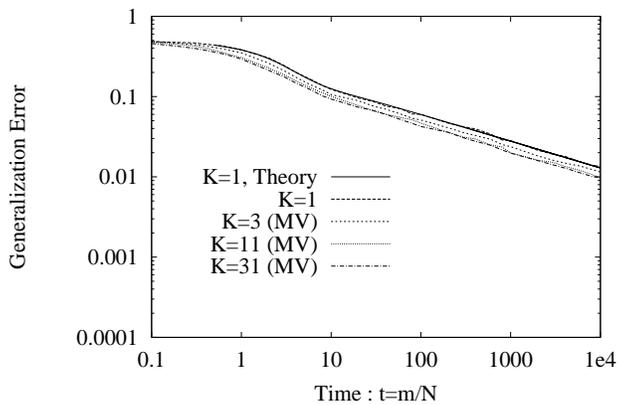}
\caption{Asymptotic behavior of generalization error
of majority vote in perceptron learning.
Computer simulations, except for the solid line.
Asymptotic order of ensemble learning is the same as
that at $K=1$.}
\label{fig:Pasysim}
\end{center}
\end{figure}

\begin{figure}[htbp]
\begin{center}
\includegraphics[width=\graphsize\linewidth,keepaspectratio]{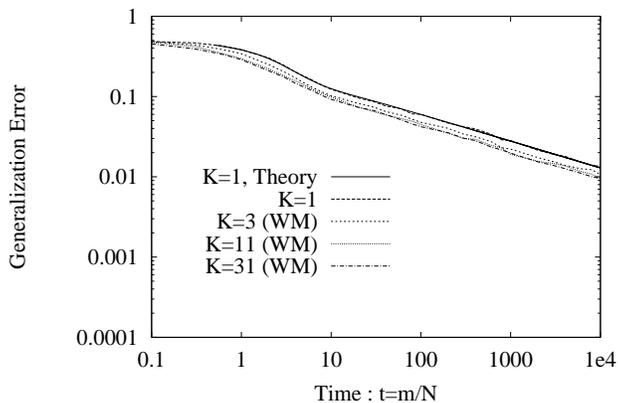}
\caption{Asymptotic behavior of generalization error
of weight mean in perceptron learning.
Computer simulations, except for the solid line.
Asymptotic order of ensemble learning is the same as
that at $K=1$.}
\label{fig:PasysimWM}
\end{center}
\end{figure}

Figures \ref{fig:Pasysim}--\ref{fig:PasysimWM}
show the results of computer simulations 
where $N=10^3$, $K=1,3,11,31$
until $t=10^4$ in order to investigate asymptotic behaviors
of generalization errors.
Effect of ensemble is maintained asymptotically.
Asymptotic behavior of generalization error
in perceptron learning
in the case of the number $K$ of students at unity
is $O(t^{-\frac{1}{3}})$\cite{NishimoriE}.
Asymptotic orders of the generalization error
in the case of ensemble learning are considered equal to
those of $K=1$, since properties of $K=3,11,31$ are
parallel to those of $K=1$ in these figures.

\begin{figure}[htbp]
\begin{center}
\includegraphics[width=\graphsize\linewidth,keepaspectratio]{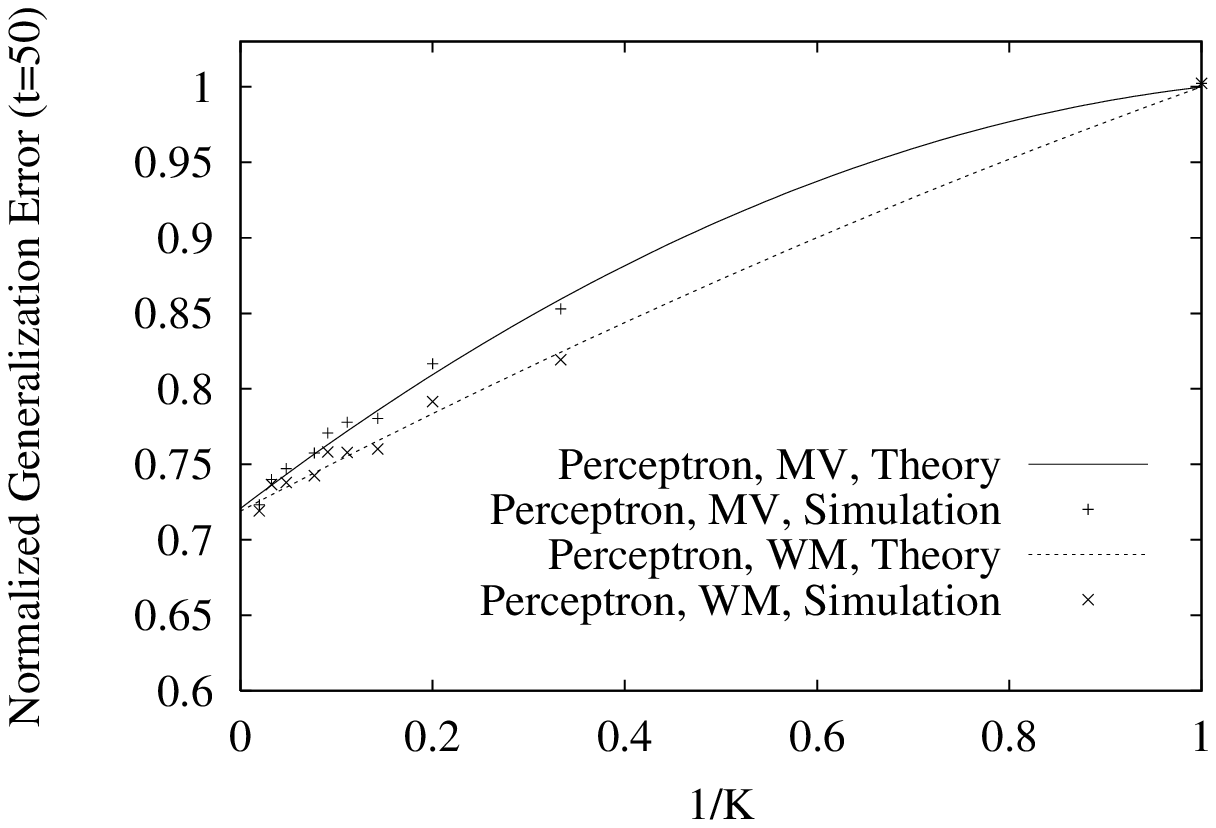}
\caption{Relationship between $K$ and effect of 
ensemble in perceptron learning.
Ensemble generalization error $\epsilon_g$ for a large $K$ limit
is about 0.72 times 
that of $K=1$.}
\label{fig:Pasy}
\end{center}
\end{figure}

To clarify the relationship between
$K$ and the effect of ensemble, 
we have obtained theoretical ensemble generalization
errors for various values of $K$. 
In the same manner as Hebbian learning,
Figure \ref{fig:Pasy} shows the results
obtained by the Metropolis method 
using the values of $R$ and $q$ calculated
numerically for perceptron learning
and Eqs. (\ref{eqn:a})--(\ref{eqn:P_1}).

\subsection{AdaTron learning}
The update procedure for AdaTron learning is
\begin{equation}
f(\mbox{sgn}(v),u)=-u\Theta\left(-uv\right).
\label{eqn:AdaTron}
\end{equation}

Using this expression,
$\left\langle f_k u_k\right\rangle $,
$\left\langle f_k v\right\rangle $ and
$\left\langle f_k^2\right\rangle $
in the case of AdaTron learning can be obtained as follows
by executing Eqs. (\ref{eqn:fkuk})--(\ref{eqn:fk^2})
analytically\cite{NishimoriE,Engel}.
\begin{eqnarray}
\left\langle f_k u_k\right\rangle  
&=& -2\int_0^\infty Du u^2
 H\left(\frac{Ru}{\sqrt{1-R^2}}\right)\\
&=& -\frac{1}{\pi}\mbox{cot}^{-1}\left(\frac{R}{\sqrt{1-R^2}}\right)
\nonumber \\
& & + \frac{1}{\pi}R\sqrt{1-R^2} \label{eqn:fuA}\\
\left\langle f_k v\right\rangle 
&=&
\frac{\left(1-R^2\right)^{\frac{3}{2}}}{\pi}
+R\left\langle f_k u_k\right\rangle ,\label{eqn:fvA}\\
\left\langle f_k^2\right\rangle &=&
 -\left\langle f_k u_k\right\rangle .  \label{eqn:f2A}
\end{eqnarray}

In this section, $\left\langle f_{k} u_{k'}\right\rangle $ and
$\left\langle f_{k} f_{k'}\right\rangle $ are derived.
Using Eq. (\ref{eqn:AdaTron}),
$\left\langle f_{k} u_{k'}\right\rangle $ and 
$\left\langle f_{k} f_{k'}\right\rangle $
in the case of AdaTron learning
are obtained as follows by executing 
Eqs. (\ref{eqn:fkuk'}) and
(\ref{eqn:fkfk'}) analytically.

\begin{eqnarray}
\left\langle f_{k} u_{k'}\right\rangle 
&=&-\int du_{k} du_{k'} dv p_3(u_{k},u_{k'},v)
    \Theta(-u_{k} v)u_{k} u_{k'} \nonumber \\
&=&\frac{1\!+\!q}{\pi}R\sqrt{1\!-\!R^2}
   - 2q\int_0^\infty Dv\int_{\frac{Rv}{\sqrt{1\!-\!R^2}}}^\infty Dx x^2
   \label{eqn:faubA} \\
\left\langle f_{k} f_{k'}\right\rangle 
&=& \int\!dvdu_{k} u_{k} du_{k'} u_{k'} p_3(u_{k},u_{k'},v),
\Theta(\!-\!u_{k} v)\Theta(\!-\!u_{k'} v) \nonumber \\
&=&
 \frac{(1\!-\!q)^2\left(1\!+\!q\!-\!2R^2\right)}
 {2\pi \left(1\!-\!R^2\right)^{\frac{3}{2}}}
 \left(\sqrt{\frac{(1\!+\!q)\left(1\!-\!R^2\right)}{1-q}}\!-\!R\right)
+
 2(q-R^2)\int_0^\infty Dv\int_{\frac{Rv}{\sqrt{1-R^2}}}^\infty Dx x^2
 H\left(z\right)
 \nonumber \\
&-&
 \frac{2R\left(1+q-R^2\right)}{\sqrt{1-R^2}}
 \int_0^\infty Dv v\int_{\frac{Rv}{\sqrt{1-R^2}}}^\infty Dx x
 H\left(z\right)
+
 2R^2
 \int_0^\infty Dv v^2\int_{\frac{Rv}{\sqrt{1-R^2}}}^\infty Dx
 H\left(z\right), \label{eqn:fafbA}
\end{eqnarray}
where 
the definitions of $z$, $H(u)$ and $Dx$ are
Eqs. (\ref{eqn:z}), (\ref{eqn:Hu}) and (\ref{eqn:Dx}), respectively.

\begin{figure}[htbp]
\begin{center}
\includegraphics[width=\graphsize\linewidth,keepaspectratio]{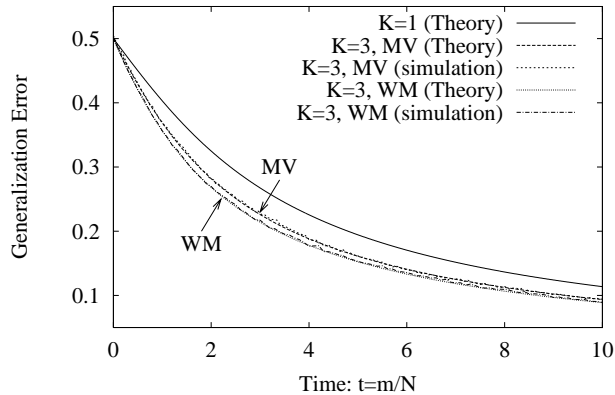}
\caption{Dynamical behaviors of ensemble generalization error $\epsilon_g$
in AdaTron learning.
Improvement of $\epsilon_g$ by increasing $K$ from 1 to 3 is 
largest of the three learning rules.}
\label{fig:AdaTron}
\end{center}
\end{figure}

In the same manner as Hebbian learning,
$R$ and $q$ have been obtained by solving 
Eqs. (\ref{eqn:dldt}), (\ref{eqn:dRdt}), (\ref{eqn:dqdt}), 
(\ref{eqn:Rqinit}), (\ref{eqn:fkuk'fkfk'}), 
(\ref{eqn:fuA})--(\ref{eqn:fafbA}) numerically.
We have obtained numerical 
ensemble generalization errors $\epsilon_g$
in the case of $K=3$ by using 
Eqs. (\ref{eqn:eg2})--(\ref{eqn:Sigma_for_eg})
and the above $R$ and $q$.
Figure \ref{fig:AdaTron} shows the results.
This figure shows that 
the ensemble generalization errors obtained 
by theoretical calculation
explain the computer simulation quantitatively.

\begin{figure}[htbp]
\begin{center}
\includegraphics[width=\graphsize\linewidth,keepaspectratio]{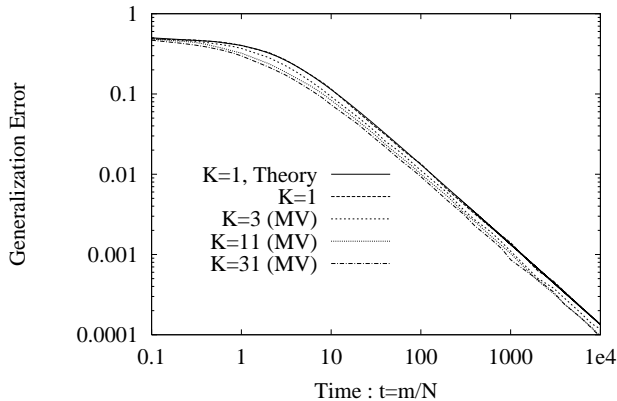}
\caption{Asymptotic behavior of generalization error
of majority vote in AdaTron learning.
Computer simulations, except for the solid line.
Asymptotic order of ensemble learning is the same as
that at $K=1$.}
\label{fig:Aasysim}
\end{center}
\end{figure}

\begin{figure}[htbp]
\begin{center}
\includegraphics[width=\graphsize\linewidth,keepaspectratio]{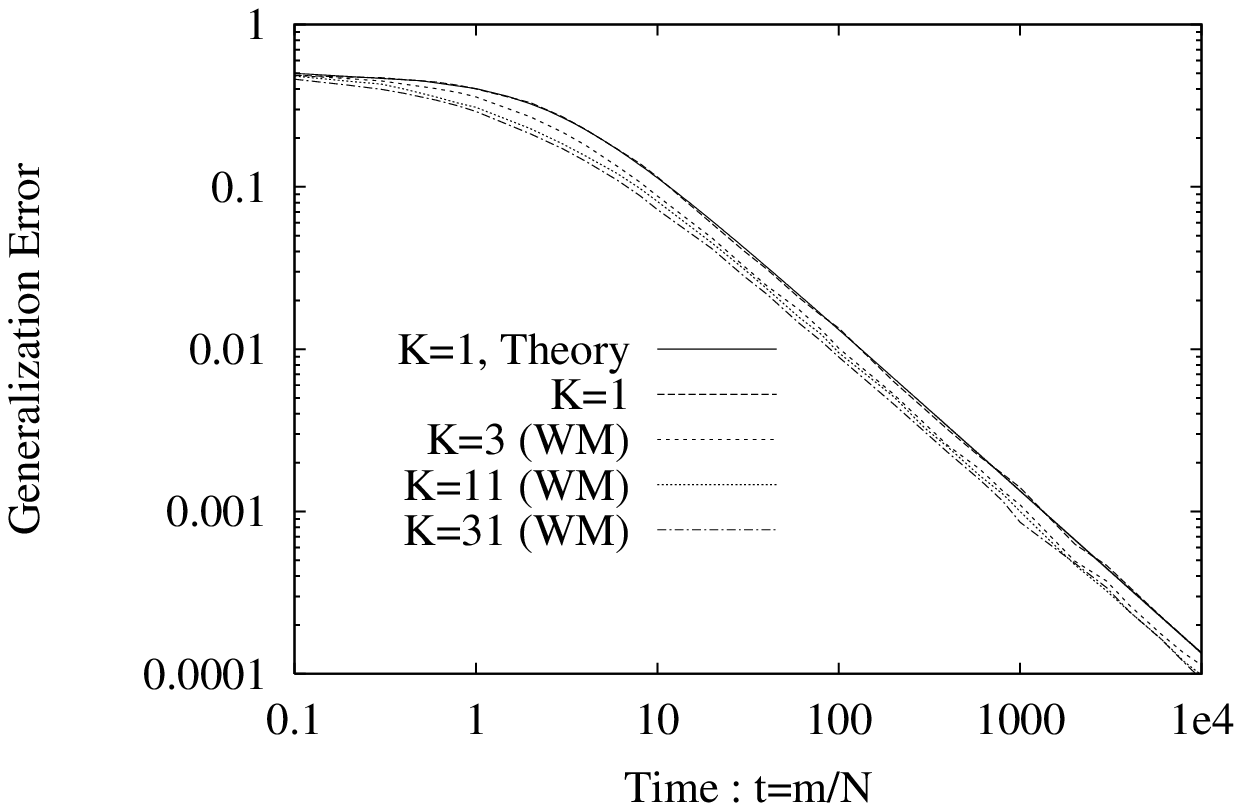}
\caption{Asymptotic behavior of generalization error
of weight mean in AdaTron learning.
Computer simulations, except for the solid line.
Asymptotic order of ensemble learning is the same as
that at $K=1$.}
\label{fig:AasysimWM}
\end{center}
\end{figure}

Figures \ref{fig:Aasysim}--\ref{fig:AasysimWM}
show the results of computer simulations 
where $N=10^3$, $K=1,3,11,31$
until $t=10^4$ in order to investigate asymptotic behaviors
of generalization errors.
Effect of ensemble is maintained asymptotically.
Asymptotic behavior of generalization error
in AdaTron learning
in the case of the number $K$ of students at unity
is $O(t^{-1})$\cite{NishimoriE,Inoue}.
Asymptotic orders of the generalization error
in the case of ensemble learning are considered equal to
those of $K=1$, since properties of $K=3,11,31$ are
parallel to those of $K=1$ in these figures.

\begin{figure}[htbp]
\begin{center}
\includegraphics[width=\graphsize\linewidth,keepaspectratio]{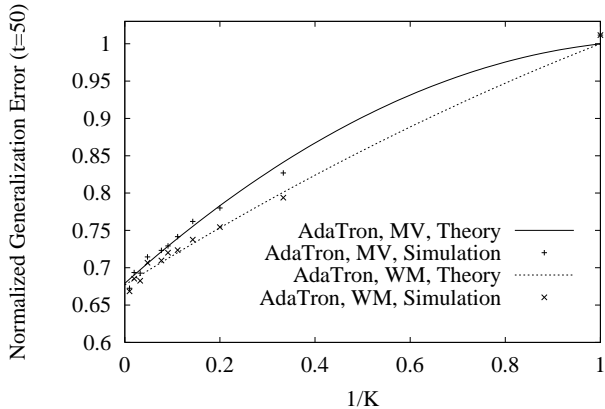}
\caption{Relationship between $K$ and effect of 
ensemble in AdaTron learning.
Ensemble generalization error $\epsilon_g$ for a large $K$ limit
is about 0.68 times 
that of $K=1$.}
\label{fig:Aasy}
\end{center}
\end{figure}

To clarify the relationship between
$K$ and the effect of ensemble, 
we have obtained theoretical ensemble generalization
errors for various values of $K$. 
In the same manner as Hebbian learning,
Figure \ref{fig:Aasy} shows the results
obtained by the Metropolis method 
using the values of $R$ and $q$ calculated
numerically for perceptron learning
and Eqs. (\ref{eqn:a})--(\ref{eqn:P_1}).

\section{DISCUSSION}
Figures \ref{fig:Hebb}，\ref{fig:Hasy}，\ref{fig:Perceptron}，
\ref{fig:Pasy}，\ref{fig:AdaTron} and \ref{fig:Aasy}
show that
the generalization errors of the three learning rules 
are all improved by ensemble learning.
However, the degree of improvement is small in 
Hebbian learning and large in AdaTron learning.
First, we discuss the reason for this difference 
in the following.

Each student moves towards teacher as learning proceeds.
Therefore, 
similarities $R_k$ and $q_{kk'}$ increase and approach unity,
leading to 
$R_k$ and $q_{kk'}$ becoming less irrelevant to each other.
For example
when $R_k=R_{k'}=1$, $q_{kk'}$ cannot be $\neq 1$ 
since a teacher $\mbox{\boldmath $B$}$,
a student $\mbox{\boldmath $J$}_k$
and another student $\mbox{\boldmath $J$}_{k'}$
have the same direction.
Thus, $R_k$ and $q_{kk'}$ are under 
a certain restraint relationship each other.
When $q_{kk'}$ is relatively smaller when compared with $R_k$,
variety among students is further maintained and 
the effect of the ensemble can be considered as large.
On the contrary,
after $q_{kk'}$ becomes unity,
a student $\mbox{\boldmath $J$}_{k}$
and another student $\mbox{\boldmath $J$}_{k'}$
are the same and there is no merit in combining them.

\begin{figure}[htbp]
\begin{center}
\includegraphics[width=\graphsize\linewidth,keepaspectratio]{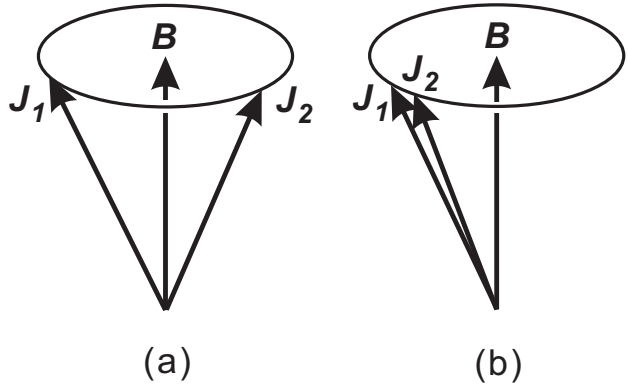}
\caption{Variety among students.}
\label{fig:effect}
\end{center}
\end{figure}

Let us explain these considerations intuitively by using 
Figure \ref{fig:effect}.
Both (a) and (b) show the relationship among two students
$\mbox{\boldmath $J$}_{1}$, $\mbox{\boldmath $J$}_{2}$
and a teacher $\mbox{\boldmath $B$}$ when 
learning has proceeded to some degree 
from the condition that the students
and the teacher have no correlation.
Then, as shown in Figure \ref{fig:effect},
students must distribute to points the same distance from
the teacher.
That is, the similarity $R_1$ of the teacher and a student
$\mbox{\boldmath $J$}_{1}$
equals the similarity $R_2$ of the teacher and a student
$\mbox{\boldmath $J$}_{2}$ in both (a) and (b).
Here, (a)
shows the case in which students are unlike each other 
--- in other words 
the variety among students is large, that is, $q$ is small.
In this case, it is obvious that a mean vector of 
$\mbox{\boldmath $J$}_{1}$ and $\mbox{\boldmath $J$}_{2}$
is closer to the teacher $\mbox{\boldmath $B$}$
than either $\mbox{\boldmath $J$}_{1}$ 
or $\mbox{\boldmath $J$}_{2}$.
Therefore, a mean vector
$\frac{1}{K} \sum_{k=1}^K \mbox{\boldmath $J$}_{k}$
of the students' connection weights
can closely approximate the connection weight vector 
$\mbox{\boldmath $B$}$ of the teacher
in cases like (a).
In addition, 
a combination method other than a mean of students,
e.g. the majority vote of students, must 
approximate the teacher better than each student 
can do alone in cases like (a).
In this case, the effect of ensemble learning is strong.
On the contrary,
Figure \ref{fig:effect}(b) shows the case in which 
students are similar to each other
--- in other words,
the variety among students is small, meaning $q$ is large.
In this case, the significance of combining
two students is small since their outputs 
are almost always the same. Therefore, 
effect of ensemble learning is small when 
$q$ is large, as in Figure \ref{fig:effect}(b).
Thus, 
the relationship between $R_k$ and $q_{kk'}$ is essential
to know in ensemble learning.

\begin{figure}[htbp]
\begin{center}
\includegraphics[width=\graphsize\linewidth,keepaspectratio]{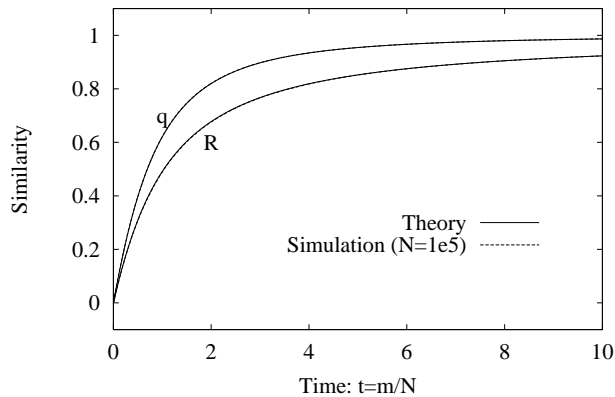}
\caption{Dynamical behaviors of $R$ and $q$ in Hebbian learning.
Here, $q$ rises more rapidly than $R$, 
which means the variety among students disappears rapidly 
in Hebbian learning.}
\label{fig:RqmergeH}
\end{center}
\end{figure}

\begin{figure}[htbp]
\begin{center}
\includegraphics[width=\graphsize\linewidth,keepaspectratio]{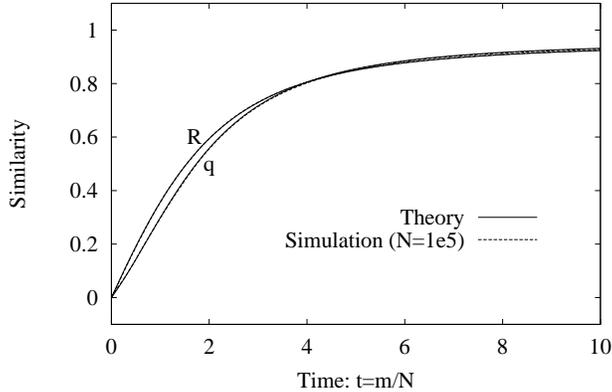}
\caption{Dynamical behaviors of $R$ and $q$ in perceptron learning.
Here, $q$ is smaller than $R$ in the early period of learning ($t<4.0$).
Perceptron learning maintains 
the variety among students for a longer time
than Hebbian learning.}
\label{fig:RqmergeP}
\end{center}
\end{figure}

\begin{figure}[htbp]
\begin{center}
\includegraphics[width=\graphsize\linewidth,keepaspectratio]{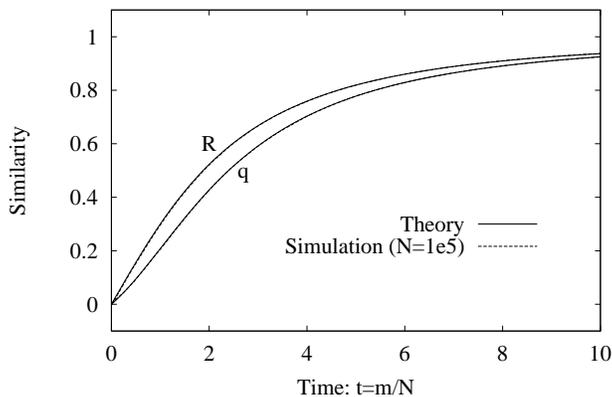}
\caption{Dynamical behaviors of $R$ and $q$ in AdaTron learning.
Here, $q$ is relatively smaller when compared with $R$
than in the cases of Hebbian learning or perceptron learning.
AdaTron learning maintains 
variety among students most out of these three learning rules.}
\label{fig:RqmergeA}
\end{center}
\end{figure}

Figure \ref{fig:RqmergeH} shows a comparison
between the theoretical results regarding the dynamical behaviors of
$R$ and $q$ of Hebbian learning, 
which are obtained by solving 
Eqs.(\ref{eqn:dldt}), (\ref{eqn:dRdt}), (\ref{eqn:dqdt}), 
(\ref{eqn:Rqinit}), (\ref{eqn:fkuk'fkfk'}), 
(\ref{eqn:fufvf2H})--(\ref{eqn:fafbH}) numerically
and by computer simulation $(N=10^5)$.
In the same manner, 
Figure \ref{fig:RqmergeP} shows a comparison
between the theoretical results regarding the dynamical behaviors of
$R$ and $q$ of perceptron learning, 
which are obtained by solving 
Eqs. (\ref{eqn:dldt}), (\ref{eqn:dRdt}), (\ref{eqn:dqdt}), 
(\ref{eqn:Rqinit}), (\ref{eqn:fkuk'fkfk'}), 
(\ref{eqn:fufvP})--(\ref{eqn:fafbP}) numerically
and by computer simulation $(N=10^5)$.
Figure \ref{fig:RqmergeA} shows a comparison
between the theoretical results regarding the dynamical behaviors of
$R$ and $q$ of AdaTron learning, 
which are obtained by solving 
Eqs. (\ref{eqn:dldt}), (\ref{eqn:dRdt}), (\ref{eqn:dqdt}), 
(\ref{eqn:Rqinit}), (\ref{eqn:fkuk'fkfk'}), 
(\ref{eqn:fuA})--(\ref{eqn:fafbA}) numerically
and by computer simulation $(N=10^5)$.
In these figures, the theoretical results and the computer
simulations closely agree with each other.
That is,
the derived theory explains the computer simulation 
quantitatively.
Figure \ref{fig:RqmergeH} shows that 
$q$ rises more rapidly than $R$ in Hebbian learning;
in other words, $q$ is relatively large when compared with $R$,
meaning the variety among students disappears rapidly 
in Hebbian learning.
Figure \ref{fig:RqmergeP} shows that 
$q$ is smaller than $R$ in the early period of learning ($t<4.0$),
which means perceptron learning maintains 
the variety among students for a longer time
than Hebbian learning.
Figure \ref{fig:RqmergeA} shows that 
$q$ is relatively smaller when compared with $R$
than in the cases of Hebbian learning and perceptron learning.
This means AdaTron learning maintains 
variety among students most out of these three learning rules.

\begin{figure}[htbp]
\begin{center}
\hspace{10mm}
\includegraphics[width=\graphsize\linewidth,keepaspectratio]{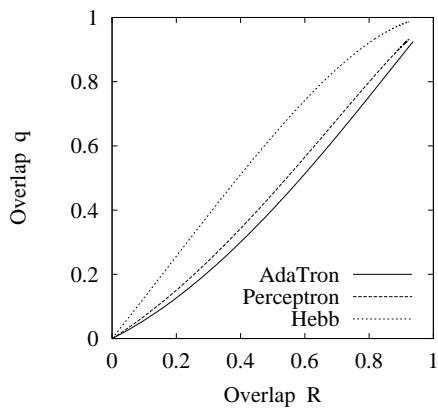}
\caption{Relationship between $R$ and $q$ (Theory).
Here, $q$ of AdaTron learning is the smallest when compared with $R$.
The rising of $q$ is the slowest and variety among students
is best maintained in AdaTron learning.}
\label{fig:Rq-APHt}
\end{center}
\end{figure}

Figures \ref{fig:RqmergeH}--\ref{fig:RqmergeA}
show that 
$q$ is relatively small when compared with $R$
in the case of AdaTron learning than in 
Hebbian learning and perceptron learning.
As described before, 
the relationship between $R$ and $q$ is essential
in ensemble learning.
To illustrate this, Figure \ref{fig:Rq-APHt} shows
the relationship more clearly by taking $R$ and $q$ as axes.
In this figure, the curve for AdaTron learning is
located in the bottom.
That is, of the three learning rules, the one
offering the smallest $q$ 
when compared with $R$ is AdaTron learning.
In other words, 
the learning rule in which the rising of $q$ 
is the slowest and the variety among students is 
maintained best is AdaTron learning.

These characteristics can be understood from the
update expression of each rule.
Equation (\ref{eqn:Hebb}) means that an update by Hebbian learning
depends on only the output $\mbox{sgn}(v)$ of a teacher.
That is, all students are updated identically at all time steps.
Therefore, the similarity of students increases rapidly in
Hebbian learning.
On the other hand,
the update by perceptron learning equals that of Hebbian learning
times $\Theta(-uv)$, as shown in Eq. (\ref{eqn:Perceptron}).
Students whose outputs are opposite to that of a teacher
change their connection weights.
At least in the initial period of learning,
students whose output is opposite to that of a teacher
and students whose output is the same as that of a teacher
both exist.
As a result, students that change their connection weights
and students who don't change their connection weights
both exist, 
leading to the fact that variety among students by perceptron learning 
is better maintained than by Hebbian learning.
The update by AdaTron learning is given in 
Eq. (\ref{eqn:AdaTron}).
This can be rewritten as $f(\mbox{sgn}(v),u)=|u|\Theta(-uv)\mbox{sgn}(v)$.
That is, the update by AdaTron learning equals 
that of perceptron learning times $|u|$, which depends on the students.
Therefore, the variety among students by AdaTron learning
is still better maintained.

In the discussion above,
the reason why the degree of improvement
by ensemble learning is small in 
Hebbian learning and large in AdaTron learning
as shown in Figures \ref{fig:Hebb}, \ref{fig:Hasy}, 
\ref{fig:Perceptron}, \ref{fig:Pasy},
\ref{fig:AdaTron} and \ref{fig:Aasy}
have been explained.
AdaTron learning originally featured the fastest asymptotic characteristic 
of the three learning rules\cite{NishimoriE}.
However, it has disadvantage that the learning is slow 
at the beginning; that is, the generalization error is 
larger than for the 
other two learning rules in the period of $t<6$.
This paper shows that 
the fastest asymptotic characteristic 
of AdaTron learning is maintained
in ensemble learning and that 
AdaTron learning
has a good affinity with ensemble learning
in regard to ``the variety among students"
and the disadvantage of the early period 
can be improved by combining it with ensemble learning.

From the perspective of the difference between 
the majority vote
and the weight mean, Figures \ref{fig:Hebb}, \ref{fig:Hasy}, 
\ref{fig:Perceptron}, \ref{fig:Pasy},
\ref{fig:AdaTron} and \ref{fig:Aasy}
show that the improvement by weight mean is larger than 
that by majority vote in all three learning rules.
Improvement in the generalization error by averaging 
connection weights of various students can be 
understood intuitively because the 
mean of students is close to that of 
the teacher in Figure \ref{fig:effect}(a).
The reason why the improvement in the majority vote is smaller 
than that in the weight mean is considered to be that
the variety among students cannot be utilized as effectively
by the majority vote as by the weight mean.
However, the majority vote can determine an ensemble output
only using outputs of students, and is easy to
implement. 
It is, therefore, significant that
the effect of an ensemble in the case of the majority vote
has been analyzed quantitatively.

Figures \ref{fig:Hasy}, \ref{fig:Pasy} and 
\ref{fig:Aasy} also show that 
the ensemble generalization errors $\epsilon_g$
by the majority vote are
larger than 
those by the weight mean in the case of $K<\infty$.
In both perceptron learning and AdaTron learning,
the relationship between $1/K$ and $\epsilon_g$
shows a straight line and an upwards-convex curve
in the case of the weight mean and the majority vote,
respectively. 
The ensemble generalization errors $\epsilon_g$ 
in the cases of 
the majority vote and the weight mean agree with each other
at a large $K$ limit.
This fact agrees with the description in \cite{Urbanczik}.
Therefore, the weight mean
is superior than the majority vote 
especially in the case of a small $K$.
Moreover, it is shown that 
$\epsilon_g$ for a large $K$ limit
compared with that of $K=1$ 
is about 0.99, 0.72 and 0.68 times
in Hebbian, perceptron and AdaTron learning, respectively.
It has been confirmed that ensemble 
has the strongest effect in AdaTron learning among 
three learning rules.

\section{CONCLUSION}
This paper discussed ensemble learning of 
$K$ nonlinear perceptrons,
which determine their outputs by sign functions
within the framework of online learning
and statistical mechanics.
One purpose of statistical learning theory
is to theoretically obtain the generalization error.
In this paper, 
we have shown that 
the ensemble generalization error can be calculated 
by using two order parameters, that is 
the similarity between the teacher and a student,
and
the similarity among students.
The differential equations that describe
the dynamical behaviors of these order parameters
have been derived in the case of general learning rules.
The concrete forms of these differential equations have been derived
analytically in the cases of three well-known rules:
Hebbian learning, perceptron learning and
AdaTron learning.
We calculated the 
ensemble generalization errors of these three rules
by using the results determined by 
solving their differential equations.
As a result, these three rules have 
different characteristics in their affinity for ensemble 
learning, that is, ``maintaining variety among students." 
The results show that 
AdaTron learning is superior to the other two
rules with respect to that affinity.

\section*{Acknowledgment}
This research was partially supported by the Ministry of Education, 
Culture, Sports, Science and Technology, Japan, 
with Grant-in-Aid for Scientific Research
13780313, 14084212, 14580438 and 15500151.


\begin{thebibliography}{99}

\bibitem{Abe}
Y. Freund and R. E. Schapire,
Journal of Japanese Society for Artificial Intelligence,
{\bf 14}(5), 771 (1999) (in Japanese, translation by N. Abe.).


\bibitem{Breiman}
Breiman, L., 
Machine Learning, {\bf 26}(2), 123 (1996).

\bibitem{Freund}
Y. Freund and R. E. Shapire,
Journal of Comp. and Sys. Sci., {\bf 55}(1), 119 (1997).

\bibitem{Hara}
K. Hara and M. Okada,
cond-mat/0402069

\bibitem{Krogh}
A. Krogh and P. Sollich,
Phys. Rev. E {\bf 55}(1), 811 (1997).

\bibitem{Urbanczik}
R. Urbanczik,
Phys. Rev. E {\bf 62}(1), 1448 (2000).

\bibitem{Hertz}
J. A. Hertz, A. Krogh and R. G. Palmer,
{\it Introduction to the Theory of Neural Computation}
(Addison-Wesley, Redwood City, CA, 1991).

\bibitem{Opper}
M. Opper and W. Kinzel,
in {\it Physics of Neural Networks III},
edited by E. Domany, J. L. van Hemmen and K. Schulten
(Springer, Berlin, 1995).

\bibitem{NishimoriE}
H. Nishimori,
{\it Statistical Physics of Spin Glasses and Information Processing: 
An Introduction}
(Oxford University Press, Oxford, 2001).

\bibitem{Anlauf}
J. K. Anlauf and M. Biehl,
Europhys. Lett. {\bf 10}(7), 687 (1989).

\bibitem{Biehl}
M. Biehl and P. Riegler,
Europhys. Lett. {\bf 28}(7), 525 (1994).

\bibitem{Inoue}
J. Inoue and H. Nishimori,
Phys. Rev. E {\bf 55}(4), 4544 (1997).

\bibitem{Saad}
D. Saad,
{\it On-line Learning in Neural Networks}
(Cambridge University Press, Cambridge, 1998).


\bibitem{NC2003-7}
S. Miyoshi, K. Hara and M. Okada,
IEICE Technical Report {\bf 103}(228), 13 (2003) (in Japanese).


\bibitem{JNNS2003}
S. Miyoshi, K. Hara and M. Okada,
Proc. Annual Conf. of Japanese Neural Network Society,
104 (2003) (in Japanese).

\bibitem{Engel}
A. Engel and C. V. Broeck,
{\it Statistical Mechanics of Learning}
(Cambridge University Press, Cambridge, 2001).

\end{thebibliography}
\end{document}